\documentclass[12pt,tightenlines,showpacs,showkeys]{revtex4-1}
\usepackage{graphicx,color,amssymb,amsmath}

\begin{document}

\title{A Study on Anisotropy in the Arrival Directions of Ultra-High-Energy
Cosmic Rays Observed by Pierre Auger Observatory}
\author{Hang Bae Kim}\email{hbkim@hanyang.ac.kr}
\affiliation{Department of Physics and The Research Institute of
Natural Science, Hanyang University, Seoul 133-791, Korea}

\begin{abstract}
We study the anisotropy in the arrival directions of PAO UHECRs,
using the point source correlational angular distance distribution.
The result shows that the anisotropy is characterized by
one prominent excess region and one void region.
The excess region is located near the Centaurus A direction,
supporting that the Centaurus A is a promising UHECR source.
The void region near the south pole direction may be used
to limit the diffuse isotropic background contribution.
\end{abstract}

\maketitle

\section{Introduction}

The cosmic ray with energy of order of $10^{20}\,{\rm eV}$ was first
reported in almost 50 years ago \cite{Linsley:1963km}.
Then, the observation of cosmic microwave background radiation
followed \cite{Penzias:1965wn}.
Soon after, Greisen \cite{Greisen:1966jv},
Zatsepin and Kuzmin \cite{Zatsepin:1966jv}
pointed out that the interaction with the cosmic microwave background
would cause the energy loss and limit the distance that
such high energy cosmic rays could travel.
This would result in the suppression in the cosmic ray energy spectrum
above the so called GZK cutoff $E_{\rm GZK}\sim4\times10^{19}\,{\rm eV}$,
if the sources are distributed over the whole universe.
If this suppression is true, as indicated by recent observations
\citep{Abraham:2008ru,Abbasi:2007sv,AbuZayyad:2012ru},
it implies that UHECR with energies above the GZK cutoff mostly come from
relatively close extragalactic sources within the GZK radius
$r_{\rm GZK}\sim100\,{\rm Mpc}$.
One consequence of this would be anisotropy in the arrival directions
of ultra-high-energy cosmic rays (UHECR),
since the matter within the GZK radius is distributed inhomogeneously
and the UHECR sources are more or less correlated with the matter distribution.
Therefore, the existence of anisotropy is an important clue
for tracing the origin of UHECR.

Searches for anisotropy in the UHECR arrival directions have been done
by using many different methods.
For the PAO data, the anisotropy manifested itself as a correlation
between the UHECR arrival directions and the locations of active galactic
nuclei (AGN) \cite{Abreu:2010ab,Cronin:2007zz}.
When we extend the correlation study to the galaxy distribution,
the conclusion is less clear than in the AGN case \cite{Koers:2008ba}.
Besides, instead of being based on the astrophysical objects,
the anisotropy search based on the auto-clustering did not provide
a strong evidence of anisotropy \cite{Abreu:2012zz}.

Surely the result of anisotropy study depends on methodology.
We need an method appropriate for the purpose we consider
and the amount of data we have.
One serious huddle in the anisotropy search
in the UHECR arrival directions is that
cosmic rays can be strongly affected by the intergalactic magnetic fields.
UHECR have such high energy that the intergalactic magnetic fields
could not completely erase the anisotropy arising from the inhomogeneous
distribution of sources.
However, the auto-clustering at small angles could be significantly weakened.
In the presence of such magnetic fields, the better way to see the clustering
due to the strong point source would be to examine the point-wise clustering
up to large angles.
The existence of a point source would manifest itself through
the local clustering of observed UHECRs about the location of the source.
This is nothing but a general principle for finding point sources
of various bands of radiation in astronomy.
In the case of UHECR, what is different from other astronomical particles
is that the spreading could be much larger and the number of data is small.

This paper reports this pointwise clustering test for the arrival
directions of UHECR observed by Pierre Auger Observatory (PAO),
aiming for both the anisotropy study and the point source search.
To see whether the local clusterings exist in the observed PAO data,
we sweep the whole sky covered by the PAO exposure by taking each point
as the reference point for the clustering of arrival directions.
We take the angular distance distribution of UHECRs about the reference point
and compare it with the one expected from the isotropic distribution.
We calculate the p-value by applying the Kuiper test on the angular
distance distribution.  The small p-value implies the departure
from isotropy about the reference point.
The departure from the isotropy can arise either from the excess
(local clustering) or the deficit (local void).
In Sec.~2, we describe the detail of statistical method we use,
including the criterion for this excess/deficit decision
based on the Kuiper test.
In Sec.~3, we present the results for the PAO UHECR arrival directions,
characterizing anisotropy by one strong excess region and one void region.
We discuss the implications of the results and conclude in Sec.~4.

\section{Anisotropy study using the single source CADD method}
\label{sec-method}

There are many ways to test the isotropy of the spherical data.
First, we may check the multipole moments.
We can also use the auto-correlation of arrival directions,
which is good for checking if there are small scale clusterings
or some regularities.

Here we use the test method suitable for
hunting for the point sources of UHECR.
We take any one point on the sphere as a reference point and examine
the distribution of arrival directions about that point.
The simple way is to look at the angular distance distribution
of arrival directions from the reference point.
In Refs.~\citep{Kim:2010zb,Kim:2012en}, we developed the simple comparison
method for the UHECR arrival direction distributions,
where the two-dimensional UHECR arrival direction distributions on the sphere
is reduced to one-dimensional probability distributions of some sort,
so that they can be compared by using the standard Kolmogorov-Smirnov (KS)
test or its variants.
For the correlation test, we adopted the reduction methods
called the correlational angular distance distribution (CADD).
The method used in this paper is simply the special case of CADD,
where the reference point is taken as if it is a single source.
When averaged over the all point sources,
this is similar to Ripley's K function,
a well-known second-order summary characteristic for the spatial pattern.
We do not take the average as we focus on the local clustering,
not on the overall clustering characteristic.
Here we present briefly the basic ideas of the reduction method
and how to calculate the p-value for the isotropy.

The correlational angular distance distribution is
the probability distribution of the angular distances of all pairs of
UHECR arrival directions and the point source directions.
As we consider the reference point as a single source,
CADD is just the probability distribution of the angular distances of
UHECR arrival directions from the reference point:
\(
\theta_{i}\equiv \cos^{-1}(\hat{\bf r}_i\cdot\hat{\bf R}),
\)
where $\hat{\bf r}_i$ ($i=1,\dots,N$) are the UHECR arrival directions and
$\hat{\bf R}$ is the reference direction.
For the comparison of CADD obtained from the data and
that from the isotropic distribution,
we can apply KS test or its variants such as Kuiper test.
In this analysis, we use Kuiper test
because it seems most suitable for our purpose and
the probability function of its statistic is available in analytic form.
The Kuiper test is based on the cumulative probability distribution (CPD),
$S_N(x)=\int^xp(x')dx'$ and the Kuiper statistic $D_{\rm K}$ is the sum
of maximum difference above and below two CPDs,
\begin{equation}
\label{KP-statistic}
D_{\rm K}=D_{\rm K+}+D_{\rm K-},
\end{equation}
where
\begin{equation}
\label{KPpm}
D_{\rm K+}=\max_{x}\left[S_{N_1}(x)-S_{N_2}(x)\right], \
D_{\rm K-}=\max_{x}\left[S_{N_2}(x)-S_{N_1}(x)\right].
\end{equation}
From the KP statistic $D_{\rm K}$,
the probability that CADD of the observed data
is obtained from the model under consideration can be estimated
using the Monte-Carlo simulations in general.
When the data in the distribution are all independently sampled,
as in our case,
the following approximate probability formula is available:
\begin{equation}
\label{KP-P-formula}
P(D_{\rm KP}|N_e) = Q_{\rm KP}([\sqrt{N_e}+0.155+0.24/\sqrt{N_e}]D_{\rm KP}),
\end{equation}
where $Q_{\rm KP}(\lambda)=2\sum_{j=1}^\infty(4j^2\lambda^2-1)
e^{-2j^2\lambda^2}$
and $N_e=N_1N_2/(N_1+N_2)$ is the effective number of data.
Now, $N_1=N_{\rm O}$, the number of observed UHECR data
and $N_2=N_{\rm S}$, the number of mock UHECR data obtained from
the isotropic distribution.
We can make the expected distribution more accurate
by increasing the number of mock data $N_{\rm S}$.
In the limit $N_{\rm S}\rightarrow\infty$, the effective number
of data is simply $N_e=N_{\rm O}$.

For a given reference point, we obtain two single point CADDs to be compared,
one from the observed PAO data and
the other expected from the isotropic distribution.
Then we calculate the p-value using the formula (\ref{KP-P-formula}).
The small p-value indicates that the distribution of arrival directions
in view of the given reference point significantly differs
from the isotropic distribution.
The departure from isotropy can be either the local excess or
the local deficit of observed UHECR around the reference point
compared to the isotropic distribution.
For small p-value, to decide whether it is the excess or the deficit,
we consider the following:
The Kuiper test uses the maximum differences of the observed distribution
above and below the expected distribution, $D_{\rm K+}$ and $D_{\rm K-}$.
Let the angular distances at which $D_{\rm K+}$ and $D_{\rm K-}$ are attained
be $\theta_+$ and $\theta_-$, respectively.
Then, the order of the angular values $\theta_+$ and $\theta_-$
can be used for this purpose.
If $\theta_+<\theta_-$, it is probably the excess.
If $\theta_+>\theta_-$, it is probably the deficit.
Of course, there can be a subtlety that small excess or deficit
very near the reference point can be missed.
But, this simple rule could catch the overall behavior correctly in most cases.

\section{Results for UHECRs observed by PAO}
\label{sec-result}

We use the UHECR data set released in 2010 by PAO \cite{Abreu:2010ab}.
It contains 69 UHECR with energy higher than $5.5\times10^{19}\,{\rm eV}$.
Their arrival directions are shown as black dots
in Fig.~2. 
The PAO site has the latitude $\lambda=-35.20^\circ$ and
the zenith angle cut of the data is $\theta_m=60^\circ$.
We use the geometric exposure function,
which is known to work well for the cosmic rays with energy higher than
the GZK cutoff.

We sweep the whole sky covered by the PAO exposure,
by taking each point as the reference point for the reduction of the arrival
direction distribution to single point CADD.
For illustration of our method, we show in Fig.~1 
the cases of two reference directions
whose p-values are smallest among excess regions and deficit regions.
The maximal excess point is ($\alpha=192.75^\circ,\,\delta=-38.77^\circ$) and
the maximal deficit point is ($\alpha=81.69^\circ,\,\delta=-67.90^\circ$),
which are marked by the $+$ symbols in Fig.~2. 
The left panels show the CADD with angular bin size of $10^\circ$.
Compared to the CADD of isotropic background (green dashed lines),
the excess and the deficit of the PAO data (black solid lines)
at small angles are clearly seen.
The right panels show the corresponding CPD,
to which we apply the Kuiper test and obtain the p-values
$P_{\rm excess,min}=2.0\times10^{-4}$ and
$P_{\rm deficit,min}=1.2\times10^{-3}$, respectively.
The vertical bars represent the sizes of $D_{{\rm K}+}$, $D_{{\rm K}-}$
and the angular distances $\theta_+$, $\theta_-$ at which they are attained.
For the excess at small angles, we have $\theta_+<\theta_-$.
For the deficit at small angles, the order is reversed.
It confirms that we can tell the excess region and the deficit region
from the order of $\theta_+$ and $\theta_-$.
However, one caveat is in order.
Because $\theta_+$ and $\theta_-$ are determined from the whole distribution
over small to large angles, sometimes small clustering at small angles
can be overlooked.
An example can be found around $\alpha=90^\circ$, $\delta=-15^\circ$ region.
It seems that there is a small clustering there, but the region is classified
into the deficit region due to the stronger deficit at middle angles.

\begin{figure}
\label{pd-PAO-CENA}
\includegraphics[width=80mm]{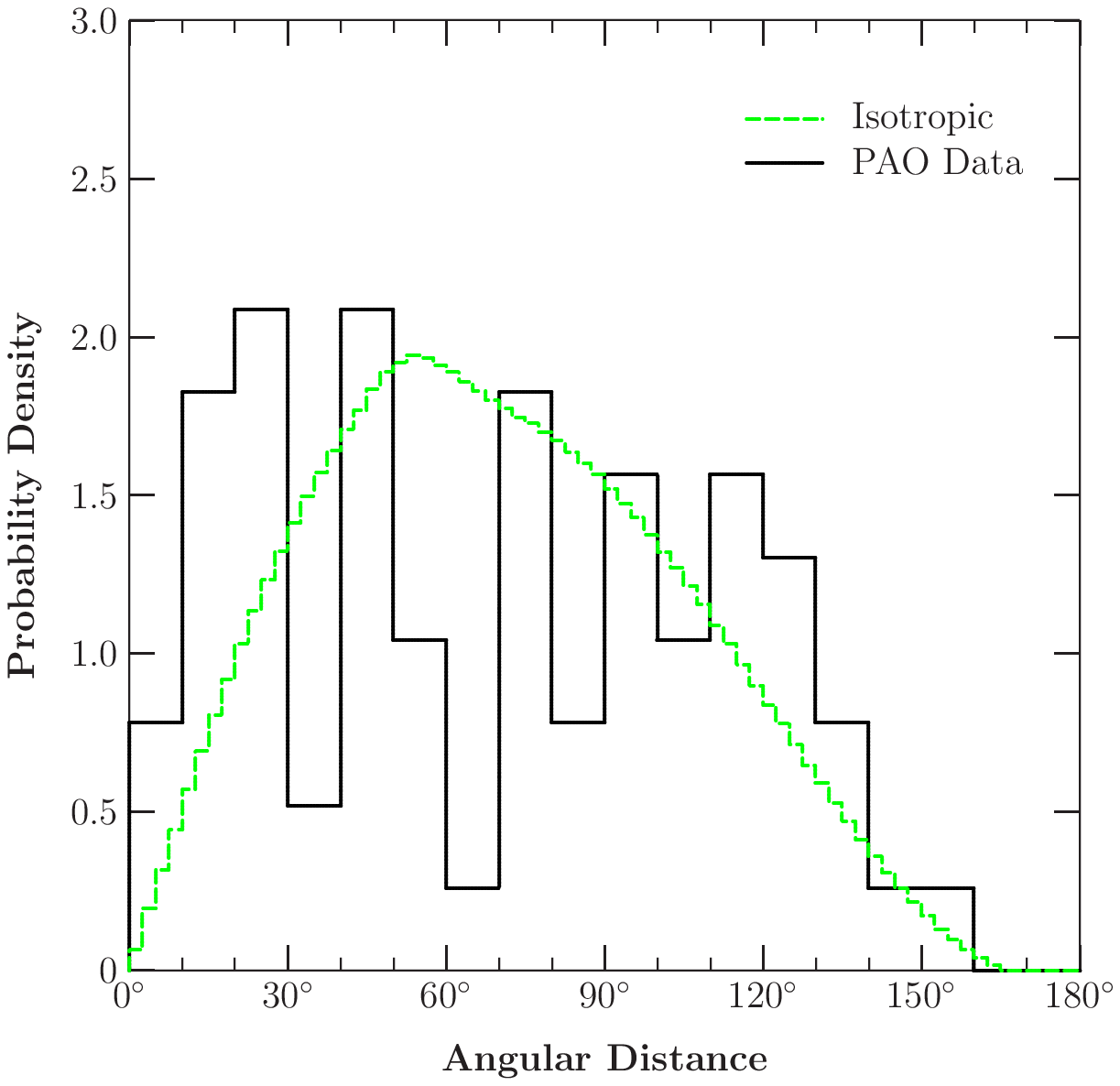}
\includegraphics[width=80mm]{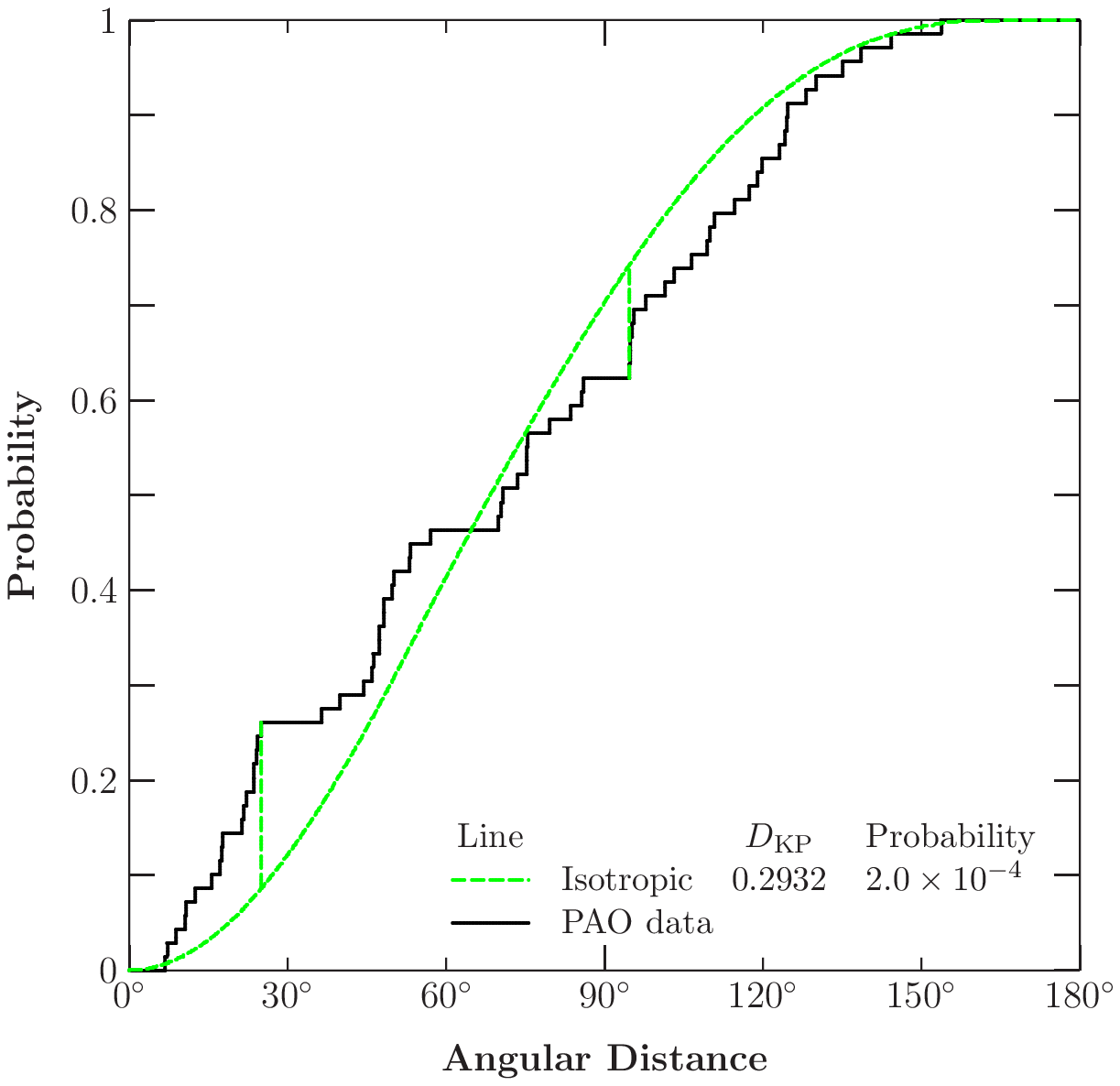}
\includegraphics[width=80mm]{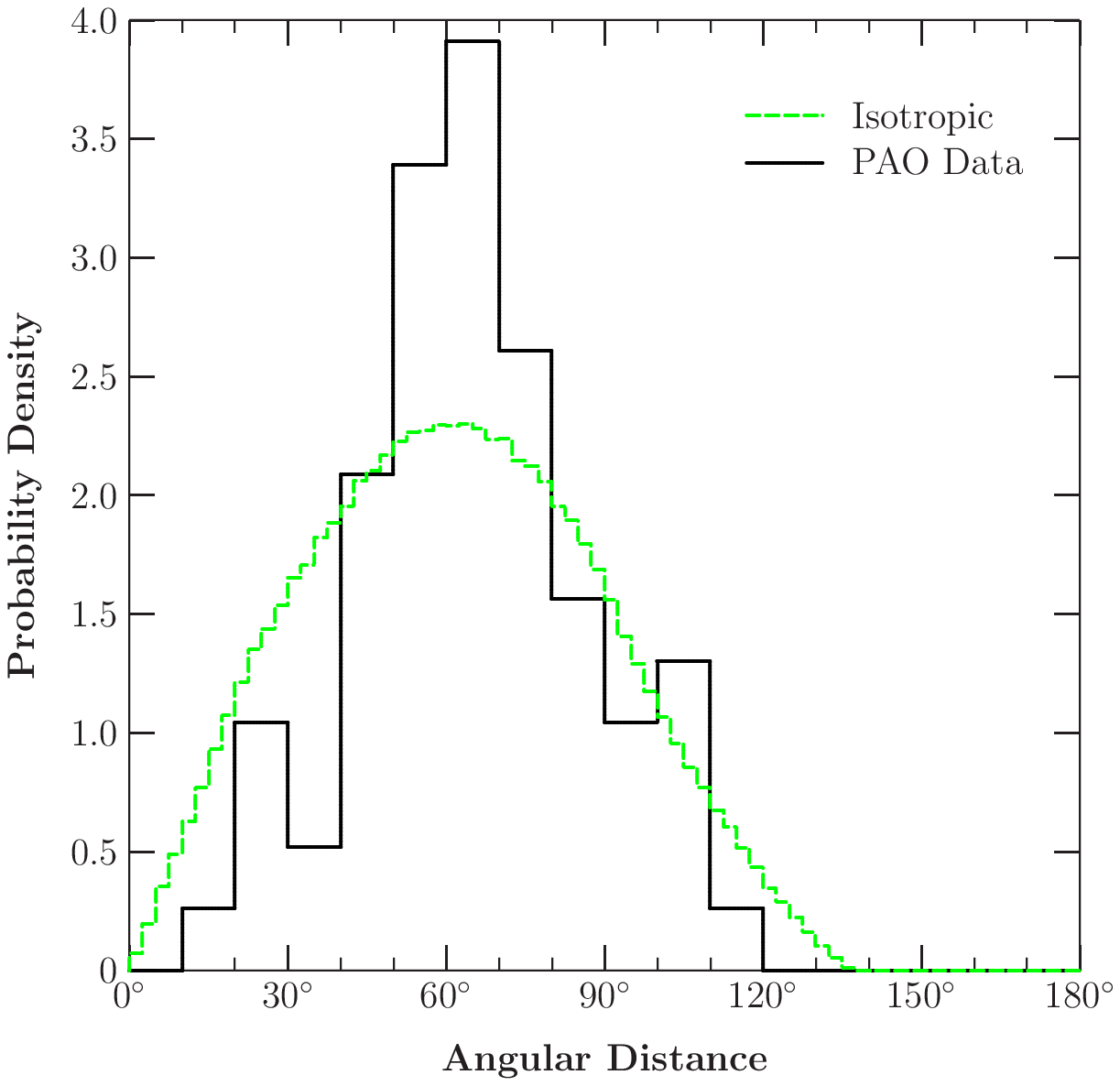}
\includegraphics[width=80mm]{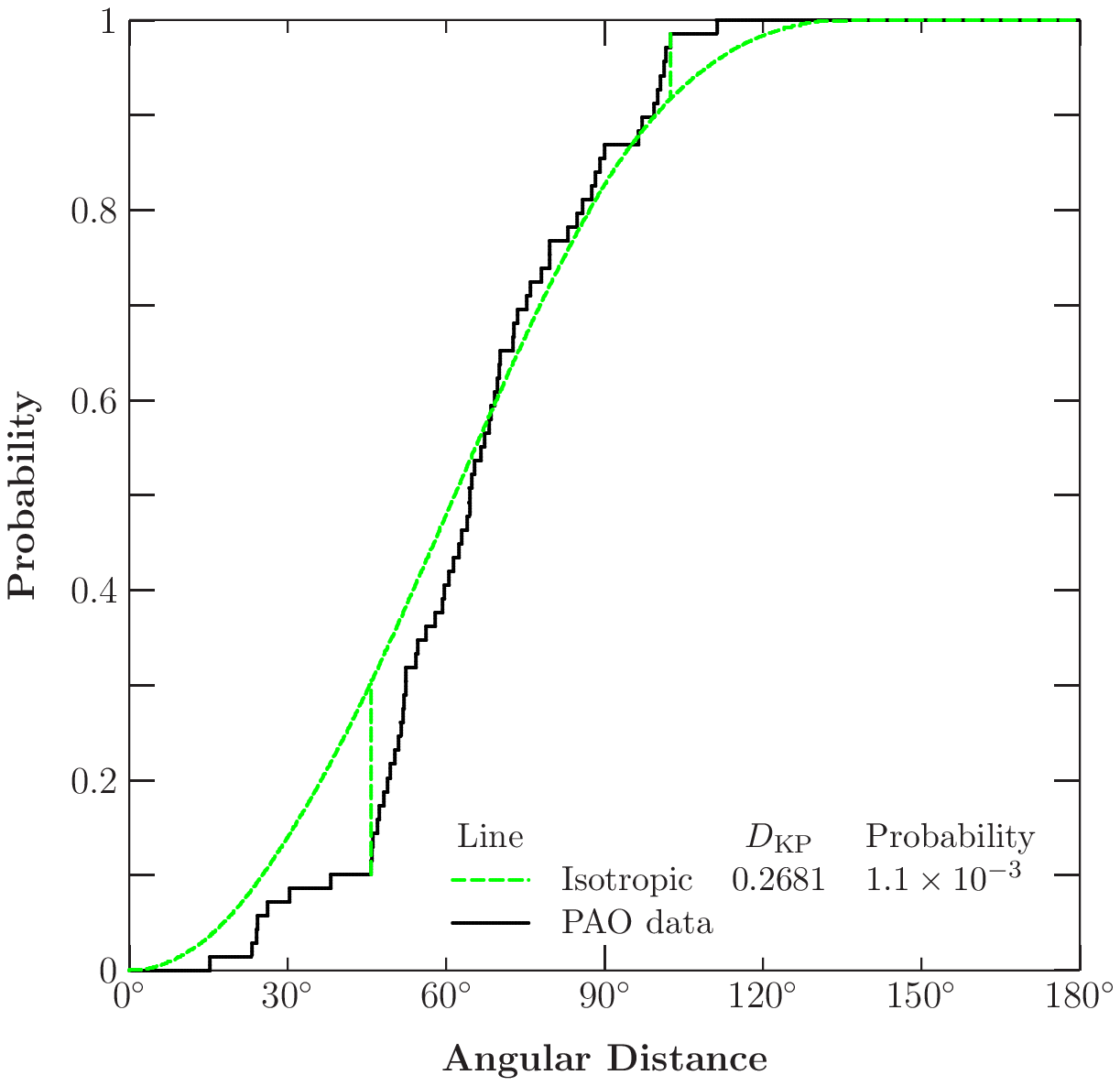}
\caption{Single point CADD and its CPD of the PAO data compared to
those of the isotropic distribution for two illustrative reference points.
The upper panels are for the maximal
excess point $(\alpha=192.75^\circ,\,\delta=-38.77^\circ)$,
and the lower panels are for the maximal deficit point
$(\alpha=81.69^\circ,\,\delta=-67.90^\circ)$.
The vertical bars in CPD represent the sizes of $D_{\rm K+}$, $D_{\rm K-}$
and the angular distances $\theta_+$, $\theta_-$
at which they are attained.}
\end{figure}

In Fig.~2, we show our main result,
the map of p-values obtained by the single source CADD method
for the PAO UHECR data.
The black dots are the arrival directions of PAO data and
graded colors represent the p-value bands
as indicated in the p-value scale bar.
Our criterion for the excess or deficit region is that
the p-value of that region is smaller than $0.0455$.
To those regions we apply the excess/deficit criterion
explained in the previous section
to further classify them into the excess or the deficit regions.
They are depicted in red color or in blue color for distinction.

\begin{figure}
\label{skymap-probability}
\centerline{\includegraphics[width=140mm]{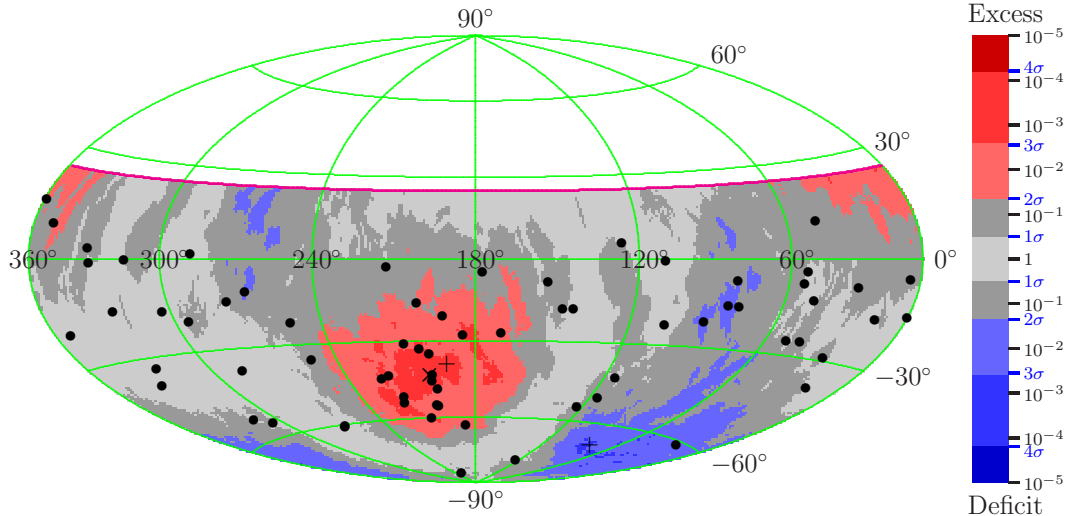}}
\caption{The p-value map for the PAO data by the single point CADD method.
The red and the blue color regions represent the excess and deficit regions
whose p-value is smaller than the $2\sigma$ value ($P<0.0455$).
The $+$ symbols mark the maximal excess and deficit points.
The $\times$ symbol marks the location of Centaurus A.}
\end{figure}

An overall feature is that we have one large excess region
and one large void region.
The large excess region is centered around the location of Centaurus A
($\alpha=201.37^\circ,\,\delta=-43.02^\circ$,
marked by the $\times$ symbol in Fig.~2).
The maximal excess point
($\alpha=192.75^\circ,\,\delta=-38.77^\circ$,
marked by the $+$ symbol in Fig.~2),
which is located inside this region,
has the p-value $P_{\rm excess,min}=2.0\times10^{-4}$.
The Centaurus A position yields the p-value
$P_{\rm excess,Cen\ A}=1.2\times10^{-3}$.
The large deficit region is located near the south pole
and it contains the maximal deficit point
($\alpha=81.69^\circ,\,\delta=-67.90^\circ$,
marked by the $+$ symbol in Fig.~2),
whose p-value is $P_{\rm deficit,min}=1.2\times10^{-3}$.
Both the excess region near Centaurus A and the deficit region
near the south pole confirms that the arrival direction distribution
of PAO UHECR is anisotropic.

\section{Conclusion}
\label{sec-conclusion}

The anisotropy in the arrival directions of UHECRs observed by PAO
was firstly noted in the correlation with AGNs.
The observed arrival directions showed more correlation with AGNs
than the isotropic distribution.
It was measured by the number of correlated events,
which lie within a fixed angular distance from AGNs.
But since the number of AGNs is larger than the number of observed
UHECRs, it is obvious that all AGNs are not the sources of UHECRs.
The method adopted in this paper is quite helpful for the search
of point sources of UHECR, since it scans the sky pointwisely
detecting the deviation from isotropy in the arrival directions.
The excess region may be an indication of UHECR source in that region.
In this regard,
the large excess region located near Centaurus A supports
the hypothesis that Centaurus A is a promising source of UHECR
\cite{Abreu:2010ab,Glushkov:2012cd}.
This fact can be used to infer the fraction of Centaurus A contribution
to the whole observed UHECRs and to estimate the size of intergalactic
magnetic fields in the vicinity of Centaurus A
\cite{Kim:2012rp}.

The distinguishing feature in our results of anisotropy study is
the identification of a void region near the south pole.
The important implication of the existence of void region
in the arrival directions of observed UHECRs is that
it limits the number of UHECRs contributed by the isotropic background,
which is presumably considered to be the
contribution from the sources outside of the GZK radius.
A detailed study on this limit is in progress.

In conclusion,
we adopted the point source correlational angular distance distrusting
to study the anisotropy in the arrival directions
of ultra-high energy cosmic rays.
Our method reveals that anisotropy in the arrival directions of
UHECR observed by PAO is characterized by
one prominent excess region and one void region,
which confirms that the arrival direction distribution is anisotropic
in the sense that it is hard to obtain from the isotropic distribution.
The excess region is located near the Centaurus A direction
supports that the Centaurus A is a promising UHECR source.
The void region near the south pole direction may be used to
limit the diffuse isotropic background contribution,
that is, the contribution from outside of the GZK radius.

\section*{ACKNOWLEDGMENT}

This research was supported by Basic Science Research Program
through the National Research Foundation (NRF) funded by
the Ministry of Education, Science and Technology
(2012R1A1A2008381).

\end{document}